\documentclass[runningheads]{llncs}
\usepackage{graphicx}
\usepackage{float}
\usepackage{multirow}
\usepackage{unicode-math}
\usepackage{hyperref}
\usepackage{booktabs} % For better table formatting

\begin{document}
\title{An Empirical Study on User Profile Analysis and SEO Performance: A Case of Taiwan Cultural Memory Bank 2.0}
\titlerunning{User Profile Analysis and SEO Performance}
\author{Mei-Yun Hsu \inst{1}\orcidID{0000-0002-4496-953X}, I-Hsien Ting\inst{2}\orcidID{0000-0002-6587-2438}, Yun-Hsiu Liu\inst{3}\orcidID{0009-0003-6606-3776} and Kazunori Minetaki\inst{4}\orcidID{0000-0003-4023-068X}}
\authorrunning{Hsu et al.}
\institute{National Museum of Taiwan, History, Taiwan\\
\email{myhsu@nmth.gov.tw}\\
\and
National University of Kaohsiung, Taiwan\\
\email{iting@nuk.edu.tw}\\
\and
National University of Kaohsiung, Taiwan\\
\email{yvonneliu0616@gmail.com}\\
\and
Kindai University, Japan\\
\email{kminetaki@bus.kindai.ac.jp}\\
}
\maketitle              % typeset the header of the contribution
\begin{abstract}

Taiwan Cultural Memory Bank 2.0 is an online curation platform that invites the public to become curators, fostering diverse perspectives on Taiwan’s society, humanities, natural landscapes, and daily life. Built on a material bank concept, the platform encourages users to co-create and curate their own works using shared resources or self-uploaded materials. At its core, the system follows a “collect, store, access, and reuse” model, supporting dynamic engagement with over three million cultural memory items from Taiwan. Users can search, browse, explore stories, and engage in creative applications and collaborative productions. Understanding user profiles is crucial for enhancing website service quality, particularly within the framework of the Visitor Relationship Management (VRM) model. This study conducts an empirical analysis of user profiles on the platform, examining demographic characteristics, browsing behaviors, and engagement patterns. Additionally, the research evaluates the platform’s SEO performance, search visibility, and organic traffic effectiveness. Based on the findings, this study provides strategic recommendations for optimizing website management, improving user experience, and leveraging social media for enhanced digital outreach. The insights gained contribute to the broader discussion on digital cultural platforms and their role in audience engagement, online visibility, and networked communication.

\end{abstract}
\section{Introduction}

Taiwan Cultural Memory Bank 2.0 \href{https://tcmb.culture.tw/}{https://tcmb.culture.tw/} is a digital platform dedicated to preserving and revitalizing Taiwan's rich cultural heritage. It serves as an online curation environment where users are empowered to become curators themselves, offering personal and collective perspectives on Taiwan’s society, humanities, natural landscapes, and everyday life. The platform is structured around a "material bank" concept, encouraging not only the use of existing resources but also the contribution of new materials. Through this, users can engage in co-creation, crafting their own curated works from a combination of shared and self-uploaded content, transforming the site into a participatory and evolving cultural archive.

\begin{figure}[H]
\centering
\includegraphics[width=0.7\textwidth]{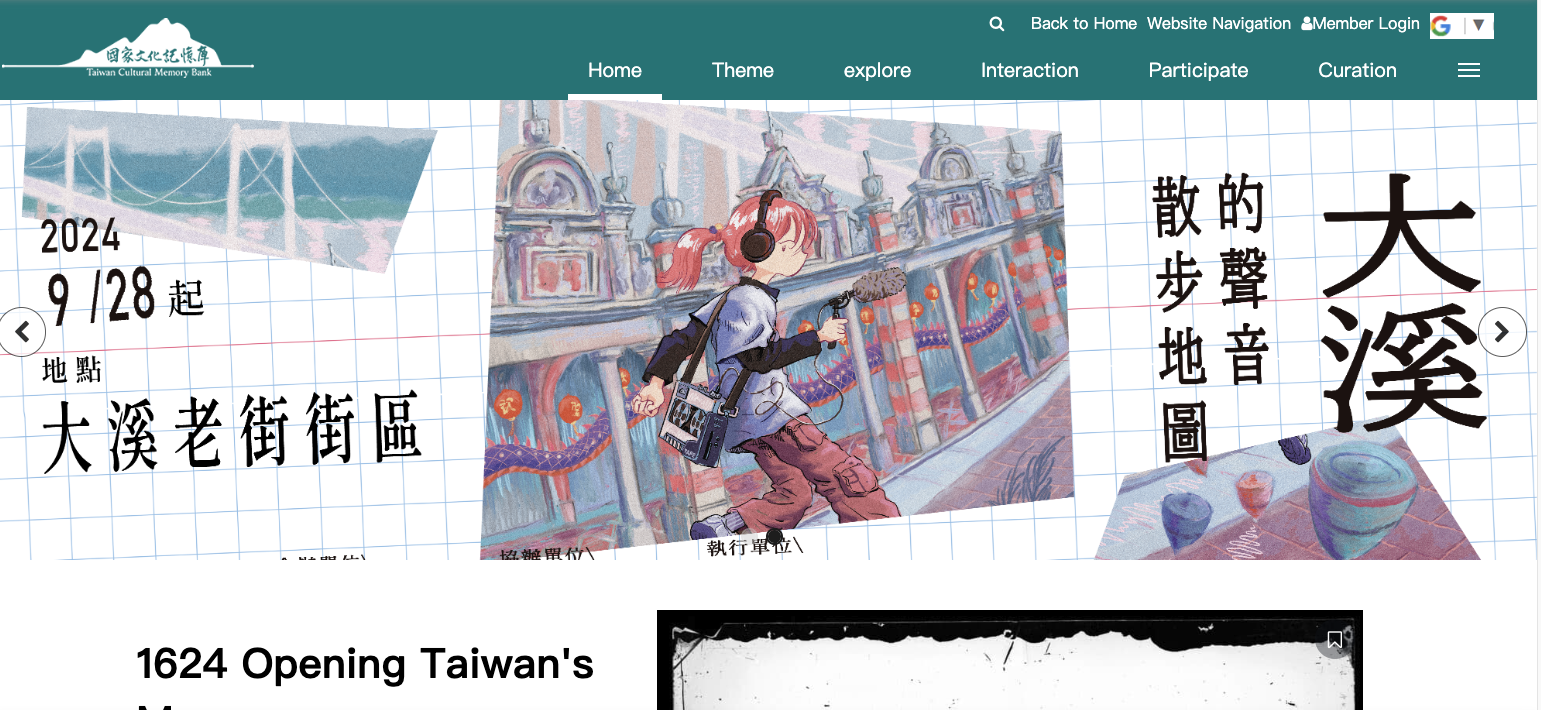}
\caption{The Website of Taiwan Cultural Memory Bank 2.0}
\label{fig1}
\end{figure}

From a system design perspective, the platform operates along four core functions: collect, store, access, and reuse. This structure ensures that cultural materials are not only preserved but also readily available for reinterpretation and creative deployment. Whether users are searching for specific themes, browsing through multimedia resources, exploring contextualized cultural stories, or assembling personalized exhibitions, the website facilitates an interactive and user-driven approach to cultural engagement. The system's flexibility and openness aim to bridge professional cultural institutions and the public, fostering a sense of shared stewardship over Taiwan’s cultural memory.

Currently, the platform houses more than three million cultural memory items from across Taiwan, making it one of the most extensive repositories of its kind. These items include images, documents, videos, oral histories, and more—each contributing to a diverse and layered representation of Taiwan’s cultural landscape. Users are invited to search, explore, and engage with these materials through story-based browsing and thematic exploration. In addition to individual use, the platform promotes creative applications and collaborative projects, supporting educational, artistic, and community-driven endeavors \cite{3}.

To better understand and enhance the effectiveness of Taiwan Cultural Memory Bank 2.0, this study focuses on analyzing user profiles and platform performance \cite{10}. Using the Visitor Relationship Management (VRM) framework, the research investigates demographic characteristics, browsing patterns, and engagement behaviors \cite{b61}. It also evaluates the site's SEO performance, organic traffic, and search visibility. Furthermore, the study incorporates social media analytics and social network analysis (SNA) to assess how content is shared and circulated across platforms \cite{b37} \cite{b46}. These insights aim to inform strategic recommendations for improving user experience, optimizing content delivery, and strengthening digital outreach through networked communication.

The remainder of this paper is organized as follows. Section 2 reviews the relevant literature on Visitor Relationship Management to establish the theoretical foundation of the study. As user profiling and search engine optimization are the two primary research focuses, related studies in these areas are also discussed in this section. Section 3 presents the findings on the online user profiles of Taiwan Cultural Memory Bank 2.0, while Section 4 details the results of the SEO performance analysis. Finally, Section 5 concludes the paper and offers recommendations for the future development of the Taiwan Cultural Memory Bank, as well as suggestions for further research.

\section{Literature Review}
\subsection{Visitor Relationship Management}

\begin{figure}[H]
\centering
\includegraphics[width=0.7\textwidth]{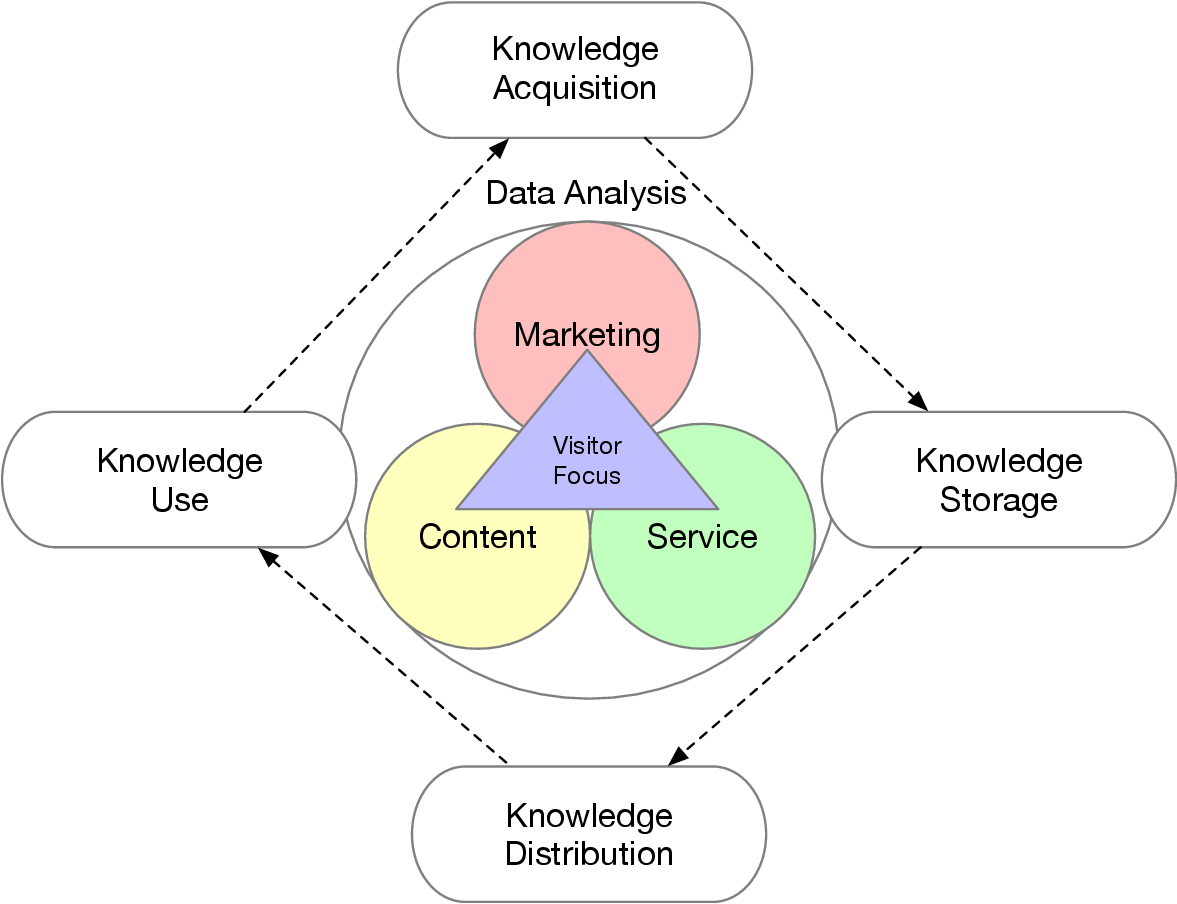}
\caption{The Visitor Relationship Management Model. \cite{b51}} \label{fig2}
\end{figure}

Visitor Relationship Management (VRM) in museums increasingly integrates concepts from customer relationship management (CRM) and knowledge management (KM) to enhance visitor engagement and institutional effectiveness \cite{9} \cite{2}. By adopting CRM principles, museums aim to better understand and respond to the needs, interests, and behaviors of their visitors \cite{6}. This approach is supported by a knowledge management framework, which provides a systematic model for managing and leveraging visitor data \cite{11}. As shown in Figure \ref{fig2}, the knowledge management cycle—comprising knowledge acquisition, storage, distribution, and use—serves as a foundation for informed decision-making across marketing, service, and content development within the museum. VRM thus becomes a strategic tool for building long-term visitor relationships and delivering more personalized cultural experiences \cite{b51}.

The integration of data collection and analysis is central to this model, enabling museums to refine their marketing strategies, enhance service quality, and develop more targeted and relevant content \cite{b1}. Through the cyclical flow of knowledge—beginning with acquisition from visitor interactions, followed by structured storage, internal distribution, and practical application—museums can adapt to visitor needs in a more agile and data-informed manner \cite{14}. At the heart of this model lies "visitor focus," which aligns marketing, content, and service in a triadic relationship designed to maximize engagement and satisfaction \cite{b63}. This knowledge-driven approach not only improves operational efficiency but also fosters deeper, more meaningful connections between museums and their audiences.

\subsection{User Profile}

A user profile is a structured collection of information that represents an individual user, typically encompassing both demographic data (such as age, gender, location) and behavioral data (such as preferences, interests, and interaction history). These profiles are foundational in digital systems where personalization, recommendation, and user-centric design are critical. In online environments, user profiles serve as a core element for tailoring content, improving user experience, and facilitating targeted communication. As digital platforms grow increasingly complex and user-driven, the role of user profiling becomes more significant in supporting data-informed strategies \cite{8}.

Understanding online user behavior is one of the most important applications of user profiling. This typically involves the analysis of user interactions captured through browsing histories, system logs, clickstreams, and other digital footprints. By mining and interpreting such data, researchers and practitioners can uncover patterns in user navigation, content consumption, and decision-making processes. These insights are particularly valuable for improving interface design, content structuring, and marketing effectiveness. Log data, in particular, provides a rich and often underutilized source for behavioral analysis, enabling dynamic and real-time understanding of user preferences \cite{b61}.

A wide range of analytical methods and tools are available to support the creation and analysis of user profiles. Demographic analysis helps segment users by background characteristics, while log analysis focuses on user interactions over time. Social network analysis can reveal relationships and influence patterns within digital communities, and data visualization facilitates intuitive interpretation of complex behavioral trends \cite{15}. Tools such as Gephi \cite{b57}, Google Analytics, and platform-specific insights tools from Facebook and YouTube allow researchers and marketers to efficiently gather, process, and present user data \cite{2b}. These methods and tools collectively enhance the capacity to build comprehensive, actionable user profiles in various digital contexts.

\subsection{SEO}
Search Engine Optimization (SEO) refers to the strategic process of enhancing a website's visibility and ranking within search engine results by understanding and aligning with the algorithms that govern search engine behavior. By optimizing various elements of a website—such as content structure, metadata, keywords, and backlinks—SEO aims to improve the site's relevance and authority in the eyes of search engines. The ultimate goal is to increase organic traffic, improve search performance, and attract new users who are actively seeking related information. In addition, SEO practices often include diagnostic insights and recommendations for website improvement, thereby contributing to overall web usability and content quality \cite{14b}.

However, the growing importance of SEO has also led to the emergence of automated techniques, including the use of bots and scripts designed to manipulate search engine rankings. These artificial methods can mimic user interactions or generate fake backlinks to boost visibility, raising concerns about the integrity of search results. As a result, recent research has focused on detecting and mitigating the impact of such bots to preserve fair competition and maintain the credibility of search engine ecosystems. The challenge lies in distinguishing between legitimate optimization efforts and deceptive practices, which continues to be an evolving area of interest in both academic and industry settings \cite{b60}.

\section{Online User Profile of Taiwan Cultural Memory Bank 2.0}

To better understand online user behavior, we use Google Analytics as our primary tool. In addition to data from the Taiwan Cultural Memory Bank 2.0, we also collect data from the Taiwan Cultural Memory Bank 1.0. The analysis covers the period from November 29, 2022, to November 2023.

\begin{figure}[H]
\centering
\includegraphics[width=1\textwidth]{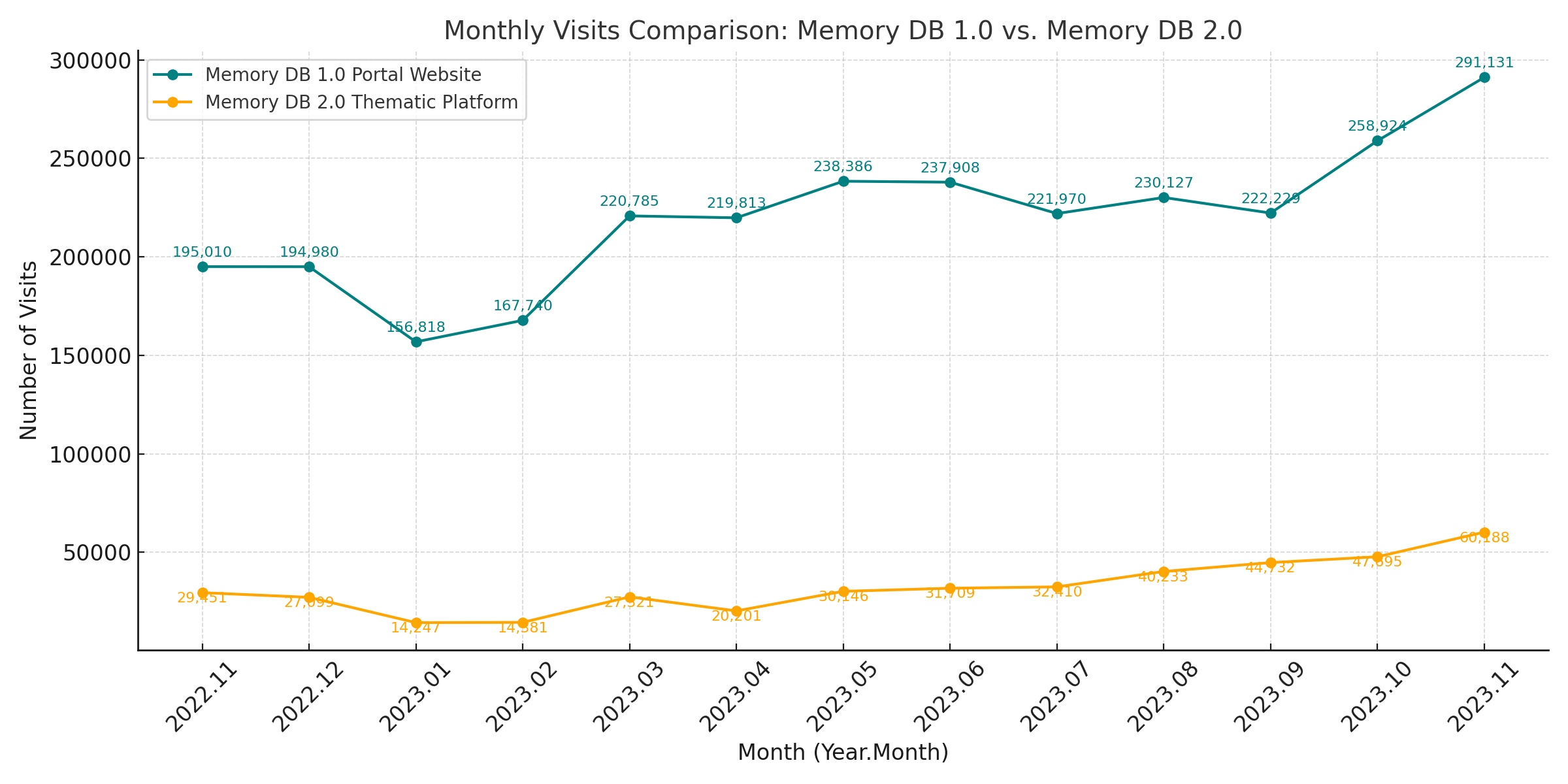}
\caption{Monthly Visits Comparison: Memory Database 1.0 vs. Memory Database 2.0} \label{fig3}
\end{figure}

Since its launch on October 17, 2020, the Memory Bank 1.0 portal has been in operation for nearly two years and has accumulated a substantial user base. Consequently, its webpage views are significantly higher compared to those of the Memory Bank 2.0 platform. Although the Memory Bank 2.0 thematic platform initially exhibited lower traffic, the project team began publishing monthly feature articles starting in May 2023. These curated topics, tailored to each month, aim to attract public interest and are promoted through various channels associated with the Memory Bank. As a result, a noticeable upward trend in page views has been observed, as illustrated in Figure \ref{fig3}.

\begin{table}[h!]
\label{table1}
\centering
\caption{User Engagement Metrics for Memory Bank Platforms}
\begin{tabular}{|l|c|c|}
\hline
\textbf{Metric} & \textbf{Memory Bank 1.0} & \textbf{Memory Bank 2.0} \\
\hline
Average Time on Site & 1 minute 9 seconds & 1 minute 56 seconds \\
\hline
Average Pages Viewed & 1.9 pages & 7.02 pages \\
\hline
Bounce Rate & 66\% & 35.2\% \\
\hline
\end{tabular}
\smallskip
\end{table}

Furthermore, in terms of basic user browsing behavior, the Taiwan Cultural Memory Bank 2.0 thematic platform outperforms the 1.0 portal website across all key engagement metrics, including average time on site, average pages viewed, and bounce rate, as shown in Table 1. These differences suggest that the higher bounce rate and lower engagement on the 1.0 portal may be attributed to system performance issues, which likely limited the depth of user interaction. In contrast, when the 2.0 platform was launched in 2023, it introduced enhancements in search functionality, more structured and thematic content, and improved UI/UX design. These optimizations have contributed to greater user engagement and increased platform stickiness. Table \ref{table1} shows the details of user engagement metrics for memory bank platforms.

The Memory Bank 1.0 portal website began to show a notable upward trend in user growth starting in September 2023. This increase is primarily attributed to recent government-funded initiatives, such as community development and Living Aesthetics Center subsidy programs, which required participating teams to upload one to three pieces of content to the “Go Create” platform as a condition for project completion. As a result, many grant recipients registered for member accounts in order to create and upload content or stories, leading to a significant rise in new users.

\begin{figure}[H]
\centering
\includegraphics[width=1\textwidth]{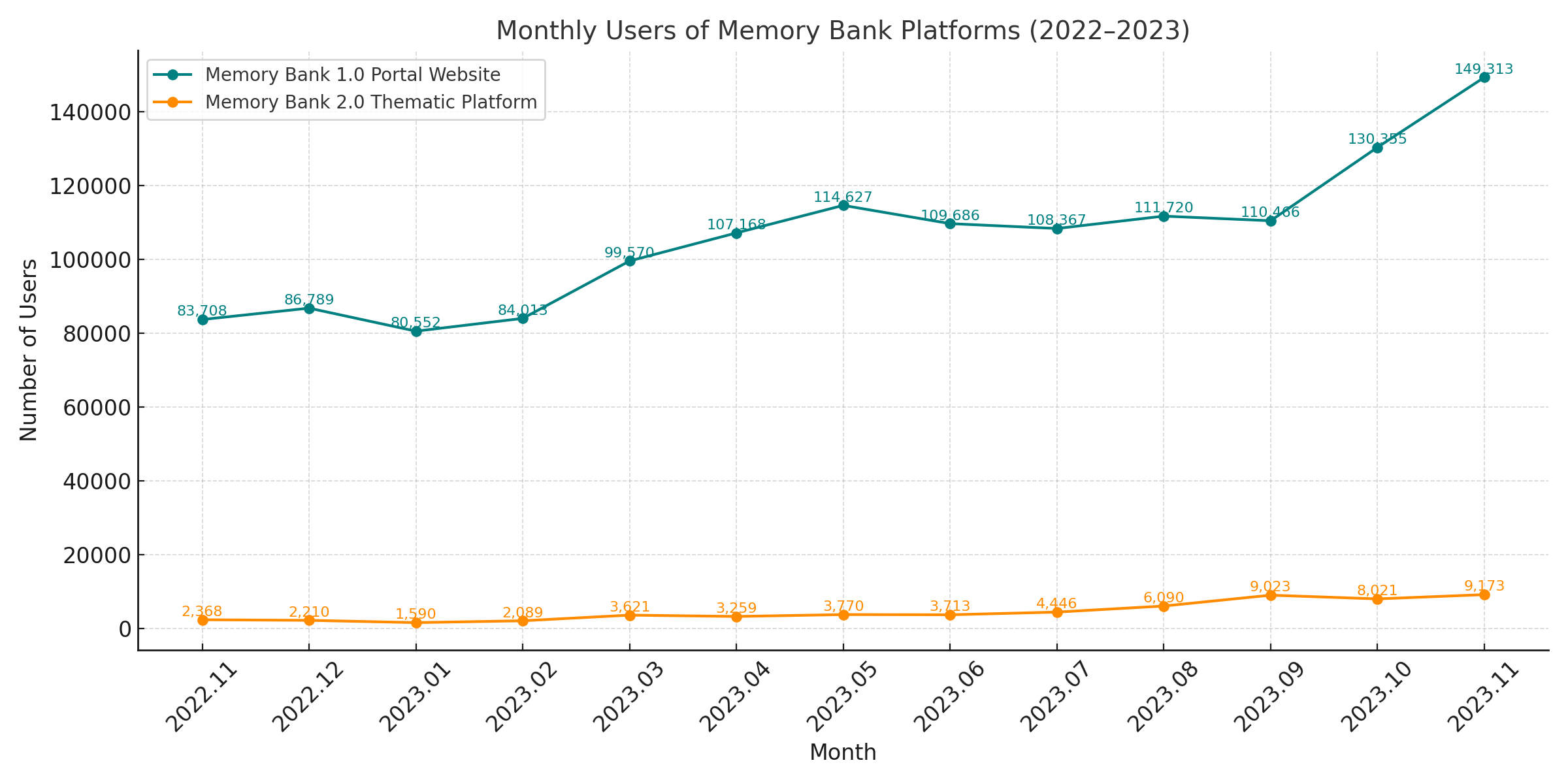}
\caption{Growth trend of new users on memory bank-related platforms} \label{fig4}
\end{figure}

\begin{table}[htbp]
\label{table2}
\centering
\caption{Top 10 Countries by Number of New Users on Memory Bank 2.0 in September, 2023}
\begin{tabular}{cccccccc}
\toprule
Rank & Country      & Number of New Users & & Rank & Country      & Number of New Users \\
\midrule
1    & Taiwan       & 7,815               & & 6    & China        & 29                  \\
2    & Singapore    & 842                 & & 7    & Malaysia     & 20                  \\
3    & United States& 103                 & & 8    & Canada       & 14                  \\
4    & Hong Kong    & 52                  & & 9    & South Korea  & 11                  \\
5    & Japan        & 49                  & & 10   & Sweden       & 9                   \\
\bottomrule
\end{tabular}
\vspace{0.2cm}
\end{table}

Similarly, the Memory Bank 2.0 thematic platform has demonstrated a steady monthly growth trend. Notably, there was a substantial increase in new users in September, with 9,023 newly registered users—an evident jump compared to the previous month. Further investigation revealed that a large portion of these users originated from Singapore, as illustrated in Figure \ref{fig4} and Table \ref{table2}, with the highest spike occurring on September 19. This suggests the possibility that a specific course or activity in Singapore was promoting the National Cultural Memory Bank. Additionally, November also saw a significant increase in new users, coinciding with the Memory Bank 2.0 "Island Translation Station" series of exchange events, which likely contributed to the surge in new user engagement.

\begin{figure}[H]
\centering
\includegraphics[width=0.7\textwidth]{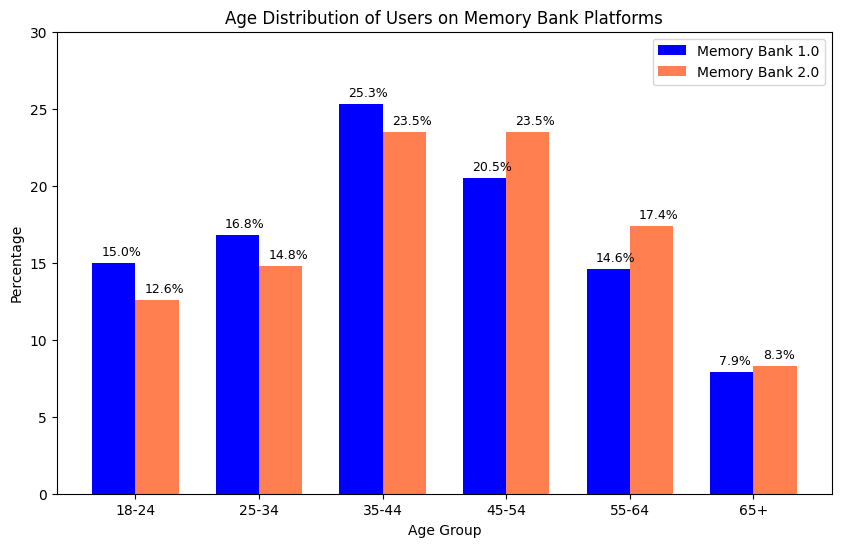}
\caption{Age Distribution of Users on Memory Bank Platforms} \label{fig5}
\end{figure}

\begin{figure}[H]
\centering
\includegraphics[width=0.7\textwidth]{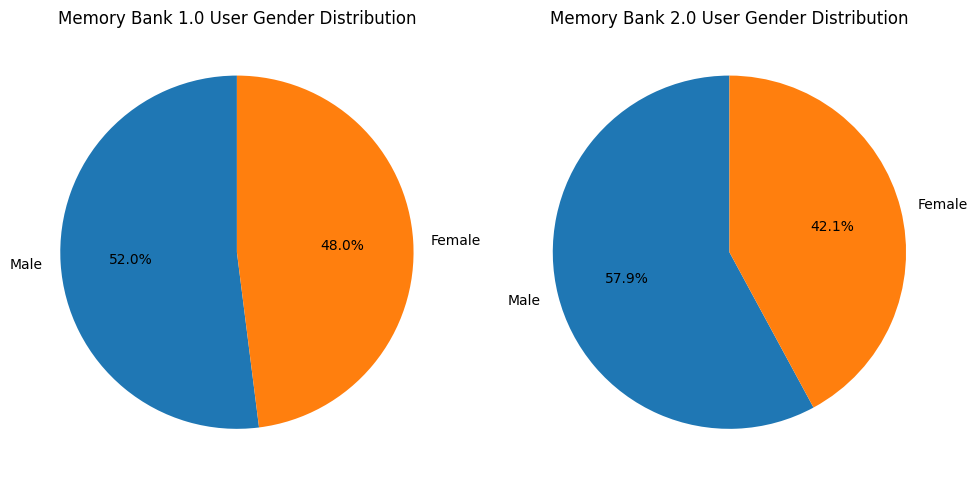}
\caption{Memory Bank 1.0 and 2.0 User Gender Distribution} \label{fig6}
\end{figure}

Figures \ref{fig5} and \ref{fig6} present the age and gender distributions within the user profiles. Notably, the age distribution of users on both the Memory Bank 1.0 portal and the Memory Bank 2.0 thematic platform demonstrates a convergent trend, predominated by users in the 35 to 54 age bracket, representing the prime working-age demographic. This age group typically corresponds to individuals in mid-career stages, exhibiting a higher demand for digital resources, which may reflect the platform's significance in supporting professional development and lifelong learning. Concurrently, the gender distribution across both platforms also reveals a consistent pattern, characterized by a majority of female users. This gender skew may be attributed to the specific nature of the content or services offered by the platforms; for instance, if the platforms focus on areas such as health, education, or social networking, these domains generally attract greater female engagement.

\section{SEO Performance of Taiwan Cultural Memory Bank 2.0}

After gaining a comprehensive understanding of the user profiles, we implemented a series of SEO strategies tailored for the Taiwan Culture Memory Bank 2.0 website. These efforts included optimizing the site structure, annotating relevant keywords, and applying various SEO treatments to enhance the website’s visibility and accessibility. Following these modifications, we collected user interaction and traffic data spanning from February 2023 to November 2024. This dataset was then analyzed using Google Analytics to evaluate user behavior, traffic sources, and overall site performance, providing valuable insights for further optimization and strategic planning.

\begin{figure}[H]
\centering
\includegraphics[width=1\textwidth]{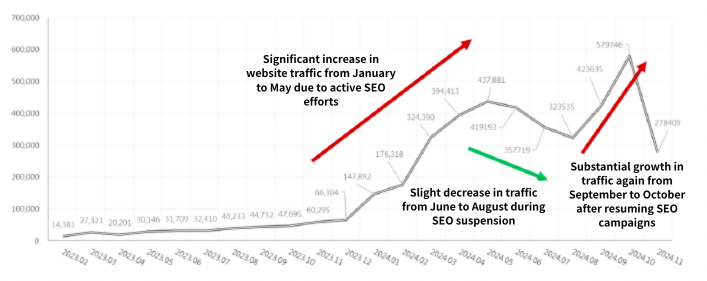}
\caption{Platform Traffic Trends and SEO Performance} \label{fig7}
\end{figure}

As shown in Figure \ref{fig7}, the Taiwan Culture Memory Bank 2.0 platform experienced a significant increase in page views beginning in January 2024, reaching a peak in May. During this period, the platform recorded an increase of approximately 1.48 million page views. This growth can be primarily attributed to the effectiveness of the implemented SEO strategies, which successfully enhanced the platform’s visibility and attracted a substantial number of users. However, from June to August, a transition period in SEO deployment led to a temporary decline in website exposure, resulting in a decrease in both new visitors and overall page views. Compared to the average monthly page views of approximately 410,000 in April and May, the numbers dropped to around 360,000 per month during the transitional phase. Subsequently, in September and October, the reimplementation of SEO strategies contributed to a significant rebound in traffic, with page views rising to approximately 500,000. Furthermore, the decline observed in November, as shown in Figure \ref{fig7}, is due to data being collected only up to November 15. Based on current trends, it is estimated that the total page views for November would exceed 500,000.

\begin{figure}[H]
\centering
\includegraphics[width=1\textwidth]{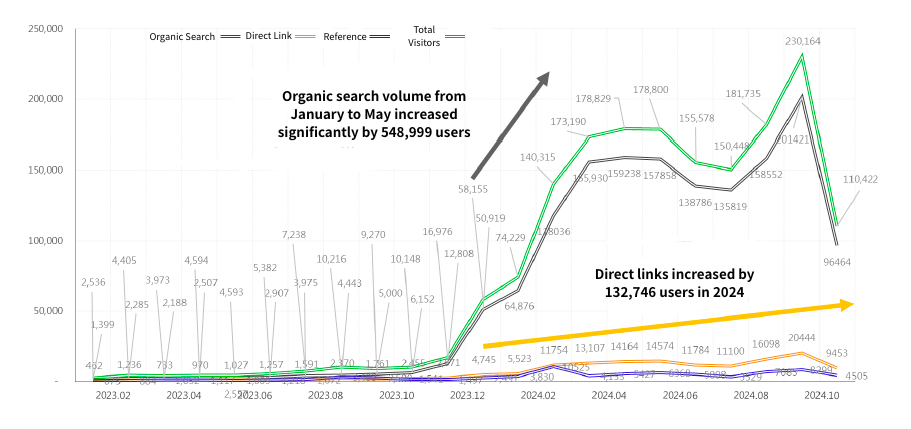}
\caption{SEO Performance of Visitor Source Analysis} \label{fig8}
\end{figure}

As shown in Figure \ref{fig8}, the majority of user traffic during the observed period originated from organic search. Following the implementation of SEO strategies, there was a noticeable and steady increase in organic search traffic starting in January 2024, with growth accelerating over time. However, a temporary suspension of SEO efforts between June and August led to a decline in organic search visits. In addition to boosting organic search traffic, SEO also contributed to an increase in direct traffic to the platform. Specifically, the average number of direct visitors rose from approximately 1,500 in 2023 to around 11,428 in 2024. These findings indicate that SEO optimization not only played a significant role in attracting new visitors but also effectively encouraged return visits from existing users. This suggests that SEO has a dual impact: it drives new traffic while also enhancing user engagement and loyalty, thereby increasing the number of dedicated platform users.

\begin{figure}[H]
\centering
\includegraphics[width=1\textwidth]{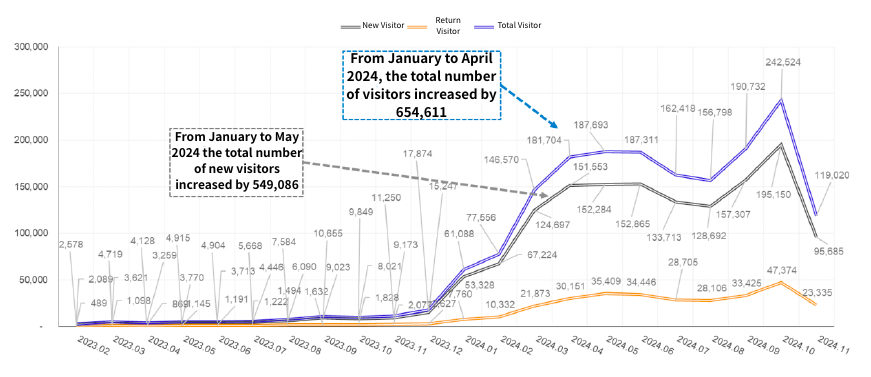}
\caption{SEO Performance of New and Return Visitors} \label{fig9}
\end{figure}

As illustrated in Figure \ref{fig9}, the number of visitors to the platform increased significantly from January 2024, reaching a peak in May. During this period, the total number of visitors grew by approximately 650,000, with the majority of this growth driven by an increase in new visitors—an estimated 540,000. This trend highlights the strong effectiveness of the SEO strategies in attracting new users to the platform. However, between June and August 2024, a transitional phase in SEO deployment resulted in a noticeable decline in new visitor traffic. Specifically, the average number of new visitors per month decreased from approximately 150,000 in April and May to around 130,000. Following the reimplementation of SEO efforts in September and October, the number of new visitors rebounded significantly, reaching approximately 190,000.

\begin{figure}[H]
\centering
\includegraphics[width=0.8\textwidth]{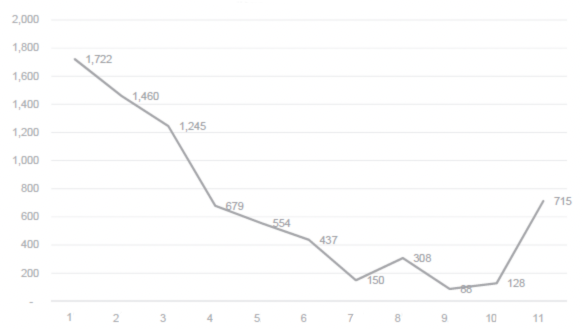}
\caption{The Increasing of Facebook Page Followers} \label{fig10}
\end{figure}

In addition to the analysis conducted using Google Analytics, we also examined the performance of the Taiwan Culture Memory Bank 2.0 Facebook fan page. The primary objective was to assess whether the SEO efforts, beyond enhancing the visibility of the official website, also had a measurable impact on the growth and engagement of the social media presence. As illustrated in Figure \ref{fig10}, the number of followers on the Facebook fan page demonstrated steady growth throughout 2024. By November 24, 2024, the page had successfully reached its target of 33,000 followers, representing an increase of nearly 8,000 followers over the course of the year. This suggests that SEO may have had a positive spillover effect on the platform’s broader digital ecosystem.

\begin{figure}[H]
\centering
\includegraphics[width=0.81\textwidth]{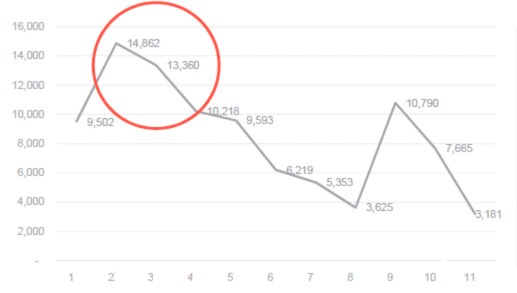}
\caption{The Number of Facebook Page Interactions} \label{fig11}
\end{figure}

Figure \ref{fig11} presents the growth in user interactions on the Taiwan Culture Memory Bank 2.0 Facebook fan page, including metrics such as post likes, comments, and shares. The period from February to March 2024 saw the highest level of engagement throughout the year, driven by content aligned with seasonal events such as the Lantern Festival and International Women’s Day. These thematic posts successfully captured audience interest and encouraged active participation. The notable increase in interactions during this period suggests that the SEO efforts not only enhanced website traffic but also contributed to improved visibility and engagement on social media platforms, demonstrating the broader effectiveness of the overall digital strategy.

\section{Conclusion and Suggestions}
In conclusion, this study highlights the critical interplay between user profile analysis, SEO optimization, and Visitor Relationship Management (VRM) in enhancing the performance of the Taiwan Cultural Memory Bank 2.0. By leveraging Google Analytics and strategic SEO interventions—such as restructuring the website, refining keyword annotations, and monitoring user behaviors—we observed substantial improvements in organic traffic, user engagement, and retention rates. These outcomes align with the core principles of VRM, which emphasize understanding and responding to visitor needs through data-informed decision-making. The platform’s improved usability and visibility reflect how VRM, when integrated with SEO strategies, can foster more personalized and engaging user experiences that support both discovery and long-term interaction.

Moreover, the study demonstrates that the impact of SEO extends beyond the official website to the platform’s social media presence. The observed growth in Facebook followers and interactions, particularly during culturally relevant periods, suggests a strong synergy between content strategy and digital outreach. These results underscore the importance of adopting a comprehensive VRM framework—one that not only captures demographic and behavioral data but also utilizes it to shape marketing efforts, optimize user engagement pathways, and strengthen audience loyalty. The findings provide a robust model for how cultural institutions can enhance their digital platforms by combining user-centered design, SEO performance, and relationship-driven strategies.

About future suggestion, Taiwan Cultural Memory Bank 2.0 is still in the stage of cultivating user engagement. To support this, SEO campaigns should be strategically scheduled to ensure continuity, helping to build user habits and loyalty. Given the clear impact of SEO on traffic growth, consistent investment and regular updates to keyword strategies based on market trends are essential. For seasonal and high-interest topics (e.g., Qixi Festival, Ghost Month), early content planning and SEO optimization, combined with social media activities, can further maximize visibility during peak periods.

Future efforts should also focus on engaging different user segments—such as lifestyle explorers, educational promoters, and creative storytellers—by offering targeted thematic content. While returning visitor rates are stable, deeper engagement can be encouraged through membership programs, personalized content, and timely updates. Tools like email notifications with tailored article suggestions can help turn new users into loyal ones and build a sustainable, active user community.


\begin{thebibliography}{11}

\bibitem{b1}
J. Anitha, I-Hsien Ting, S. Akila Agnes, S. Immanuel Alex Pandian, R.V. Belfin, Chapter 3 - Social media data analytics using feature engineering, Editor(s): J. Dinesh Peter, Steven L. Fernandes,
In Advances in ubiquitous sensing applications for healthcare,
Systems Simulation and Modeling for Cloud Computing and Big Data Applications, Academic Press, 2020, Pages 29-59, https://doi.org/10.1016/B978-0-12-819779-0.00003-4.


\bibitem{2}
Bitgood, S. \& Shettel, H. 1996., An Overview of Visitor Studies. Journal of Museum Education, 21(3):6-10.

\bibitem{2b}
Cutroni, J. (2010). Google Analytics: understanding visitor behavior. " O'Reilly Media, Inc.".

\bibitem{3}
Falk, 1998., A Framework for Diversifying Museum Audience., Museum 
News,77(5):36-39,61.

\bibitem{6}
Hooper-Greenhill, E. (1994). Museums and Their Visitors (1st ed.). Routledge. \doi{10.4324/9780203415160}.

\bibitem{b51}
Hsu, MY., Ting, IH. (2023). Applying a Combination Model of Knowledge Management and Visitor Relationship Management in the Study of the Visitors of Historical Museum. In: Uden, L., Ting, IH. (eds) Knowledge Management in Organisations. KMO 2023. Communications in Computer and Information Science, vol 1825. Springer, Cham. https://doi.org/10.1007/978-3-031-34045-1\_23Hsu, MY., Ting, IH. (2023). Applying a Combination Model of Knowledge Management and Visitor Relationship Management in the Study of the Visitors of Historical Museum. In: Uden, L., Ting, IH. (eds) Knowledge Management in Organisations. KMO 2023. Communications in Computer and Information Science, vol 1825. Springer, Cham. https://doi.org/10.1007/978-3-031-34045-1\_23

\bibitem{8}
E. Raad, R. Chbeir and A. Dipanda, "User Profile Matching in Social Networks," 2010 13th International Conference on Network-Based Information Systems, Takayama, Japan, 2010, pp. 297-304, doi: 10.1109/NBiS.2010.35.




\bibitem{9}
Sanchez, R (1996) Strategic Learning and Knowledge Management, Wiley, Chichester

\bibitem{b37}
Scott, J. (2000). Social Network Analysis: A Handbook. SAGE Publication.

\bibitem{10}
Sheng, C. W., Chen, M. C. (2012) A study of experience expectations of museum visitors, Tourism Management, Volume 33, Issue 1,
2012, Pages 53-60, \doi{10.1016/j.tourman.2011.01.023}.

\bibitem{11}
Siu, N.Y.M., Zhang, T.J.F., Dong, P., Kwan, H.Y. (2013). New service bonds and customer value in customer relationship management: The case of museum visitors, Tourism Management, Volume 36, 2013, Pages 293-303, \doi{10.1016/j.tourman.2012.12.001}.

\bibitem{b60}
I-Hsien Ting, Kazunori Minetaki, Mei-Yun Hsu, and Chia-Sung Yen. 2024. Applying Social Network Embedding and Word Embedding for Socialbots Detection. In Proceedings of the 2023 IEEE/ACM International Conference on Advances in Social Networks Analysis and Mining (ASONAM '23). Association for Computing Machinery, New York, NY, USA, 712–718. https://doi.org/10.1145/3625007.3627306

\bibitem{b61}
Ting, I. H. Special Issue: Applications and Management Aspects of Social Networks Research. Rev Socionetwork Strat 16, 571–572 (2022). https://doi.org/10.1007/s12626-022-00130-y

\bibitem{b63}
Ting, I. H., Zhang, Z. and Wang, L. S. L. 2016. A Study of The Effect of Spiral of Silence between Different Social Networking Platforms. In Proceedings of the The 3rd Multidisciplinary International Social Networks Conference on SocialInformatics 2016, Data Science 2016 (MISNC, SI, DS 2016). Association for Computing Machinery, New York, NY, USA, Article 55, 1–5. https://doi.org/10.1145/2955129.2955183

\bibitem{b56}
I-Hsien Ting and Chia-Sung Yen (2012). Opinion Groups Identification in Blogsphere based on the Techniques of Web and Social Networks Analysis. In The 4th International Conference on Machine Learning and Computing. Hong Kong, China.



\bibitem{b57}
Wajahat, A., Nazir, A., Akhtar, F., Qureshi, S., Razaque, F., \& Shakeel, A. (2020, January). Interactively visualize and analyze social network Gephi. In 2020 3rd international conference on computing, mathematics and engineering technologies (iCoMET) (pp. 1-9). IEEE.

\bibitem{b46}
Wasserman, B., \& Faust, K. (1994). Social Network Analysis: Methods and Applications. Cambridge University Press.

\bibitem{14}
Wielinga, B., J. Sandberg, and G. Schreiber, Methods and Techniques for Knowledge Management: What Has Knowledge Engineering to Offer? Expert Systems With Applications, 1997. 13(1): p. 73-84.

\bibitem{14b}
Yalçın, N., \& Köse, U. (2010). What is search engine optimization: SEO?. Procedia-Social and Behavioral Sciences, 9, 487-493.

\bibitem{15}
C. Ziemkiewicz, A. Ottley, R. J. Crouser, K. Chauncey, S. L. Su and R. Chang, "Understanding Visualization by Understanding Individual Users," in IEEE Computer Graphics and Applications, vol. 32, no. 6, pp. 88-94, Nov.-Dec. 2012, doi: 10.1109/MCG.2012.120.










\end{thebibliography}
%

\end{document}